\documentstyle[preprint,aps,epsf]{revtex}

\newcommand{\beq}{\begin{equation}}
\newcommand{\eeq}{\end{equation}}
\newcommand{\beqa}{\begin{eqnarray}}
\newcommand{\eeqa}{\end{eqnarray}}

\def\npb#1{Nucl.\ Phys.\ {\bf B #1}}
\def\plb#1{Phys.\ Lett.\ {\bf B #1}}
\def\prd#1{Phys.\ Rev.\ {\bf D #1}}
\def\prl#1{Phys.\ Rev.\ Lett. {\bf #1}}

\def\gsim{\ \rlap{\raise 3pt \hbox{$>$}}{\lower 3pt \hbox{$\sim$}}\ }
\def\lsim{\ \rlap{\raise 3pt \hbox{$<$}}{\lower 3pt \hbox{$\sim$}}\ }
\newcommand{\wppo}{\mbox{$\omega \to \, \pi^{0}\, \pi^{0}\,\gamma\,$}}
\newcommand{\wpp}{\mbox{$\omega \to \, \pi\, \pi\,\gamma\,$}}
\newcommand{\wppp}{\mbox{$\omega \to \, \pi^{+}\, \pi^{-}\,\gamma\,$}}
\newcommand{\wplus}{\mbox{$\omega \to \, \pi^{+}\, \pi^{-}\,\pi^{0}\,$}}
\newcommand{\Bsg}{\mbox{${B^{*}} \to \, B\, \gamma\,$}}
\newcommand{\Dsg}{\mbox{${D^{*}} \to \, D\, \gamma\,$}}

\begin{document}

%\draft

{\tighten
\preprint{\vbox{\hbox{ }}}

\title{$\omega-\rho$ Mixing and the $\omega\rightarrow\pi\pi\gamma$
Decay}

\author{Dafne Guetta\,$^{1}$ and   Paul Singer\,$^{2}$ }

\footnotetext{\footnotesize E-mail addresses:
$1$ firenze@physics.technion.ac.il \\
\,\,\,\,\,\,\,$2$ phr26ps@physics.technion.ac.il  }

\address{ \vbox{\vskip 0.truecm}
  Department of Physics,\\
Technion- Israel Institute of Technology,\\ Haifa
32000, Israel}

\maketitle

\begin{abstract}%
We reexamine the \wppo decay, adding the effect of
$\omega-\rho$ mixing to the amplitude calculated
with the aid of chiral perturbation theory and
vector meson dominance. We predict the neutral
decay to occur with a width of
$\Gamma(\wppo)=(390\pm96) {\rm eV}$ and also
analyze the effect of the $\omega-\rho$ mixing on
the $\Gamma(\wppo)/\Gamma(\wppp)$ ratio. Several
remarks on the effect of $\omega-\rho$ mixing on
certain radiative decays of vector mesons are
presented.
\end{abstract}
} % end tighten
\bigskip
\leftline{ PACS number(s):  12.39.Fe, 13.20.Jf, 12.40.Vv}

\vspace{1cm}

\noindent

The radiative decays of mesons is a subject of
continuos interest on both the experimental and
theoretical planes since the early sixties. The
main effort has been directed firstly to the
magnetic dipole transitions $V\to P\gamma,\,P\to
V\gamma,$ where $P,V$ belong to the lowest
multiplets of vector $(V)$ and pseudoscalar mesons
$(P).$ A large variety of theoretical models has
been employed to treat these transitions, like
quark models, bag models, effective Lagrangian
approaches, potential models, sum rules and other
(Refs. \cite{donnell,singer,dolinsky,barik}
provide a comprehensive and complementary list of
references). Recentely, the interest has focused
on such transitions in the sector of the heavy
mesons, i.e. \Bsg, \Dsg. Here again the models
mentioned above have been used, this time in
combination with heavy quark effective theories
(\cite{cheng,cho,paul}, see Ref.\cite{paul} for an
extensive list of references).

Another class of electromagnetic radiative decays
of vector mesons is that in which the final state
consists of more than one hadron, like \wpp
\cite{paul1}, $\rho\to\pi\pi\gamma$
\cite{paul2,renard}, $\phi\to K\bar{K}\gamma$
\cite{nussinov,close}, $\phi\to\pi\pi\gamma$
\cite{renard,achasov} and similar decays with one
$\eta-$meson in the final state
\cite{fajfer,bramon}. Although these decays have
smaller branching ratios than $V\to P\gamma$
decays, their study offers several attractive new
physics features, like the possibility of
investigating final state interactions in the
hadronic $\pi\pi$ \cite{achasov,levy,gubin,marco}
and $K\bar{K}$ \cite{nussinov,close} channels as
well as affording the application of chiral
perturbation theory for their calculation
\cite{marco,bramon1,huber}.

The original model \cite{paul1} for the \wpp decay
postulated a mechanism involving the dominance of
the intermediate vector meson contribution (VMD);
i.e., the transition occurs via
$\omega\to(\rho)\pi\to\pi\pi\gamma.$  Thus, the
basic interaction term is the Wess-Zumino anomaly
term of the chiral lagrangian \cite{wess}
proportional to the Levy-Civita antisymmetric
tensor. The interaction term of two vector mesons and one
pseudoscalar meson is then given by
\beq
{\cal L}_{V_{1}V_{2}\pi}=f_{V_{1}V_{2}\pi}
\epsilon_{\alpha\beta\gamma\delta}q_{1}^{\alpha}
\epsilon_{1}^{\beta}q_{2}^{\gamma}\epsilon_{2}^{\delta}
\eeq
where $q_{i},\epsilon_{i}$ are the respective
momenta and polarizations ($V_{i}$ may be a
photon). Using this mechanism, the Born amplitude
for \wppo decay is given by
\beqa \label{born}
A^{(B)}(\wppo) &=&
\frac{f_{\omega\rho\pi}f_{\rho\pi\gamma}}{m_{\pi}^2}
\epsilon_{\alpha\beta\gamma\delta}\epsilon_{\nu\delta\tau\psi}
p^\alpha{\epsilon^{*}}^{\beta}
(p)p_{3}^{\tau}\epsilon^{\psi}(p_{3})
\nonumber \\
& & \left[
\frac{P^{\gamma}P^{\nu}}{(P^{2}-m_{\rho}^{2}+i\Gamma_{\rho} m_{\rho})}
+\frac{Q^{\gamma}Q^{\nu}}{(Q^{2}-m_{\rho}^{2}+i\Gamma_{\rho}
m_{\rho})}\right],
\eeqa
with
\beq
P=p_{2}+p_{3},\,\,Q=p_{1}+p_{3}
\eeq
where $p_{1},p_{2}$ are pion momenta and $p_{3}$
and $p$ are the photon and $\omega$ momenta.

The decay with of \wppo is proportional to
$g^{2}_{\omega\rho\pi}$ and
$g^{2}_{\rho\pi\gamma}.$ Assuming that $\omega\to
3\pi$ proceeds via the same mechanism,
$\omega\to(\rho)\pi\to\pi\pi\pi$ \cite{gellmann}
and using the experimental input $\Gamma(\omega\to
3\pi) = (7.47\pm 0.14){\rm MeV},
\Gamma(\rho^{0}\to \pi^{0}\gamma)=(102.5\pm 25.6){\rm KeV},
\Gamma(\rho\to \pi\pi)=(150.7\pm 1.1){\rm MeV}$ \cite{pdg}
one predicts \cite{paul1,bramon}, using the
Born-term amplitude (\ref{born}),
\beq
\label{gamma}
\Gamma^{B}(\wppo)=\frac{1}{2}\Gamma^{B}(\wppp)
=(344\pm 85){\rm eV},
\eeq
where the factor $1/2$ is a result of charge
conjugation invariance to order $\alpha$
\cite{paul1} which imposes pion pairs of even
angular momentum.
In calculating (\ref{gamma}) we used in eq.(\ref{born}) a
momentum dependent width for the $\rho$-meson \cite{connell1}
\beq
\Gamma_{\rho}(q^2)=\Gamma_{\rho}
\left(\frac{q^2-4 m_{\pi}^2}{m_{\rho}^2-4 m_{\pi}^2}\right)^{3/2}
\frac{m_{\rho}}{\sqrt{q^2}}.
\eeq

If a constant $\rho$ width is used as frequently done 
(see, e.g.\cite{langacker}), a width 
of only 306 eV is obtained for \wppo  from the Born term.
 The value in (\ref{gamma})
obtained with presently known coupling constants,
updates the different older values in the existing
literature.

The branching ratio $\Gamma(\wppo)/\Gamma(\wplus)$
is independent of $g_{\omega\rho\pi}$ and is a
function of $g_{\rho\pi\gamma},\,g_{\rho\pi\pi}$
only in this model. Presently there appears to be
a discrepancy between the experimental
$\rho^{0}\to\pi^{0}\gamma$ and
$\rho^{+}\to\pi^{+}\gamma$ widths. We shall return
to this point later. In obtaining (\ref{gamma}) we
used the experimental value of
$\rho^{0}\to\pi^{0}\gamma$ for both the 
charged and the neutral \wpp  decays. The
quantities mentioned before eq.(\ref{gamma}) give
$\Gamma(\wppo)/\Gamma(\wplus)=(4.6\pm 1.2)\times
10^{-5},$ using the Born term (\ref{gamma}) only.
Recently, a new theoretical approach has been
advanced for the calculation of $V\to
P^{0}{P^{\prime}}^{0}\gamma$ decays \cite{bramon}
by using the framework of chiral perturbation
theory. In Ref.\cite{bramon} various decays of the
$V\to P P^{\prime}\gamma$ type have been
calculated at the one loop level, including both
$\pi\pi$ and $K\bar{K}$ intermediate loops. As
these authors have shown, the one-loop
contributions are finite and to this order no
counterterms are required. The calculation
\cite{bramon} has covered the
$\phi^{0}\to\pi\pi\gamma,\,K\bar{K}\gamma,\pi^{0}\eta^{0}\gamma,
\rho^{0}\to\pi^{0}\pi^{0}\gamma,\pi^{0}\eta\gamma $ and
$\omega\to\pi^{0}\pi^{0}\gamma,\pi^{0}\eta\gamma $ decays.
An improved calculation for the $\phi^{0}$ and
$\rho^{0}$ decays using unitarized chiral
amplitudes \cite{marco} leads to comparable
numerical results.

Now, in addition to the chiral loop contribution,
there is always an additional term in the decay
amplitudes given by the intermediate vector meson
dominance (VMD) mechanism
\cite{paul1,fajfer,bramon}. In the $\rho^{0}$
decays, the contribution from pion loops is
comparable to that given by VMD term while kaon
loops give a minute contribution only. In the
$\phi^{0}$ decays, the contribution from kaon
loops (pion loops are isospin forbidden) is an
order of magnitude larger than the Zweig-forbidden
VMD term \cite{bramon1}.

On the other hand, a very different and
interesting situation arises in the $\omega\to
\pi^{0} \pi^{0}\gamma$ decays. Here, as a result of
isospin invariance, only kaon loops can contribute
and the chiral amplitude $A_{\chi}$ obtained
\cite{bramon1} is very small; the VMD amplitude
$A_{\rm VMD}$ is the dominant feature and leads to
a decay width which is two orders of magnitude
larger than $A_{\chi}$ gives. Thus, for \wppo we
have the remarkable result
\beq
A(\wppo)=A_{\chi}(\wppo)+A_{\rm VMD}(\wppo)\simeq
A_{\rm VMD}(\wppo).
\eeq

This implies that the decay width for this mode
should be essentially accounted for by the
amplitude (\ref{born}). A similar situation is
encountered in the
$\omega\to\pi^{0}\eta^{0}\gamma$ decay
\cite{bramon1}; however, in view of the very small
branching ratio of $\sim 10^{-7}$ expected for it,
we shall not discuss further this mode here.

At this point we refer to the experimental
situation. For the charged mode there is only an
upper limit, ${\rm Br}(\wppp)<3.6\times 10^{-3}$
\cite{pdg}. On the other hand the neutral mode has
been detected by the GAMS-Collaboration at HEP
\cite{alde} and the branching ratio has been
measured to be ${\rm Br}(\wppo) =(7.2\pm
2.6)\times 10^{-5}.$ Using the well determined
$\omega$ full width of $(8.41\pm 0.09){\rm MeV}$
one arrives at at $\Gamma(\wppo)=(0.61\pm
0.23){\rm KeV}.$ The central value of this result
is nearly twice the VMD result of
eq.(\ref{gamma}). This lead us to reexamine the
mechanism of this decay, especially in light of
its unique position described above, which
requires $A(\wppo)\simeq A_{\rm VMD}(\wppo).$

Within the theoretical framework just described,
based on chiral perturbation theory and vector
meson dominance, there is one feature which has
been neglected so far. This is the possibility of
$\omega-\rho$ mixing \cite{connell} which, for
example, is responsible for the isospin violating
$\omega\to\pi^+\pi^-$ decay, occuring with a
branching ratio of \cite{pdg} ${\rm
Br}(\omega\to\pi^+\pi^-) = (2.21\pm 0.30) \%.$

We proceed now to investigate whether the 
$\omega-\rho$ mixing could possibly account for 
the existing discrepancy between the central values of 
the theoretical (\ref{gamma}) and experimental 
results.

The mixing between the isospin states $\rho^{(I=1)},\,
\omega^{(I=0)}$ may be described by adding to the
effective Lagrangian a term ${\cal L} = {\cal M}^2_{\rho\omega}
\omega_{\mu}\rho^{\mu},$ which leads to the physical
states
\beq \label{mix}
\rho=\rho^{(I=1)}+\epsilon \omega^{(I=0)},\,\,
\omega=\omega^{(I=0)}-\epsilon \rho^{(I=1)}
\eeq
where \cite{connell}
\beq \label{eps}
\epsilon = \frac{ {\cal M}^2_{\rho\omega}}
{m_{\omega}^2-m_{\rho}^2+i m_{\rho}\Gamma_{\rho}
-i m_{\omega}\Gamma_{\omega}}.
\eeq
Using  the experimental values for $m_{\rho},\,
m_{\omega},\,\Gamma_{\rho},\,\Gamma_{\omega}$
\cite{pdg} and ${\cal M}^2_{\rho\omega}=-(3.8\pm 0.4)
\times 10^{3}{\rm MeV}^2$ as determined from fits
to $e^+e^-\to\pi^+\pi^-$ \cite{connell}, one
obtains
\beq
\epsilon =-0.006 + i 0.036.
\eeq
The effect of the $\omega-\rho$ mixing is to add
to the Born diagram the two diagrams of Fig.1,
expressing the mixing of $\rho$ into the $\omega$
wave function  (\ref{mix}) as well as the
modification arising from mixing in the $\rho$
propagator \cite{langacker}. As a result, the full
amplitude for $\wppo$ decay is given by

\beq \label{ampmix}
A^{(\omega-\rho)}(\wppo)=
\tilde{A}(\wppo)+\epsilon A^{(B)} (\rho^0\to\pi^0\pi^0\gamma),
\eeq
where $\tilde{A}$ has the form of eq.(\ref{born})
with the $\rho$ propagator replaced by
\beq
\frac{1}{P^2-m_{\rho}^2+i m_{\rho}\Gamma_{\rho}}
\rightarrow  \frac{1}{P^2-m_{\rho}^2+i m_{\rho}\Gamma_{\rho}}+
\frac{f_{\omega\pi\gamma}}{f_{\rho\pi\gamma}}
\frac{ {\cal M}^2_{\rho\omega}}
{(P^2-m_{\rho}^2+i
m_{\rho}\Gamma_{\rho})(P^2-m_{\omega}^2+i
m_{\omega}\Gamma_{\omega})}.
\eeq
In the second term in (\ref{ampmix})
$A^{(B)}(\rho^0\to\pi^0\pi^0\gamma),$ has the same
expression as (\ref{born}), except that
$\omega,\rho$ are interchanged everywhere.
Calculating now the decay width from
(\ref{ampmix}) we find
\beq \label{res}
\Gamma^{(\omega-\rho)}(\wppo)= (363\pm 90){\rm eV}.
\eeq
Thus, $\omega-\rho$ mixing increases the \wppo
width by $5\%$ only, even less than the $12\%$ increase provided 
by the $q^2$-dependence of $\Gamma_{\rho},$ as discussed after 
eq.(\ref{gamma}).
The newly calculated
value is still about half the experimental one
\cite{pdg,alde} of $\Gamma^{\rm (exp)}(\wppo) =(610\pm
230){\rm eV}.$ We have checked the effect of using
the experimental value of $\rho^+\to\pi^+\gamma$
instead of $\rho^0\to\pi^0\gamma$ and the effect
of the relatively slight changes in the value of
 ${\cal M}^2_{\rho\omega},$ as given in the literature \cite{connell},
and we found that the result given in (\ref{res})
is practically not changed. It should also be
mentioned at this point that the uncertainty in
(\ref{res}) is mostly due to the uncertainty in
$\Gamma(\rho^0\to\pi^0\gamma).$

Since $\omega-\rho$ mixing turns out to be a small effect
here also, we combine now all the improvements on the simple
Born term of \cite{paul1}, i.e. $\omega-\rho$ mixing,
$q^2$-dependence of $\Gamma_{\rho}$ and the inclusion of the
$A_{\chi}$ term as given in eq.(4) of \cite{bramon1}.
Using the amplitude which contains all these effects we 
predict 

\beq 
\Gamma^{th}(\wppo)=(390\pm 96) {\rm eV}.
\eeq

Before concluding, we present a few remarks related to the effects
of  $\omega-\rho$ mixing on various radiative decays of vector
mesons.

The $\omega-\rho$ mixing expressed in Fig.1
affects  the neutral mode \wppo, hence the $1/2$
ratio of eq.(\ref{gamma}) between the neutral and
the charged modes will be affected as well,
becoming slightly larger. This is understandable,
since the $1/2$ factor holds to the first order in
$\alpha,$ while the amplitude (\ref{ampmix})
contains terms of order $e^3.$ At this point, it
is important to refer to the photon spectrum of the
$\wpp$ decay which, as shown in Refs.
\cite{paul1,bramon} peaks very strongly around 325
MeV. This is the typical spectrum of the direct
transition, as driven by the Born term of
(\ref{born}). There is however an additional
effect to this mode which was pointed out by
Fajfer and Oakes \cite{fajfer} and is caused by
the bremstrahlung radiation
$\omega\to\rho\to\pi^+\pi^-\gamma$ \cite{paul2},
emitted following an  $\omega-\rho$ transition. In
this case, contrary to the situation discussed in
this paper, only the \wppp decay will be affected.
The $1/2$ factor will change again , this time in
the opposite direction, becoming as small as $\sim
1/5$ \cite{fajfer}. This effect holds however,
mostly for the lower part of the photonic
spectrum, being due to the bremstrahlung radiation
of $\rho^0$ and practically dies out beyond
$E_{\gamma} \sim 250 {\rm MeV}.$ Hence both
effects can be experimentally tested if enough
events are collected to separate the spectra at
the photon energy of about 300 MeV.

A second remark refers to statements made in the
literature which attribute to $\omega-\rho$ mixing
or other isospin/$SU(3)$ breaking effects the
inducement of a very large deviation from 1 of the
ratio
$R=\Gamma(\rho^0\to\pi^0\gamma)/\Gamma(\rho^+\to\pi^+\gamma),$
predicting as much as $R=2.4$ \cite{diaz} or
$1.7\pm0.1$ \cite{hashimoto}. Unfortunately, these
are based on an erroneous formulation of vector
meson mixing. Ref\cite{connell} presents in detail
the correct treatment for this problem. Using
(\ref{mix}) and (\ref{eps}) to calculate the
effect of $\omega-\rho$ mixing on $R,$ we expect
$R=\Gamma(\rho^0\to\pi^0\gamma)/\Gamma(\rho^+\to\pi^+\gamma)=1.03$
which is consistent with the experimental figure
\cite{pdg} of $R^{\rm exp} =1.51\pm 0.54.$

We also wish to remark that we 
did not discuss here effects of final state 
interactions. These are not expected to change 
the prediction for the rate \cite{levy,marco},
although  may affect somewhat the decay spectrum.
In any case, if the discrepancy we discussed survives
after more accurate experiments, this point should 
be reexamined as well.

We summarize by stressing the importance of a good
measurement of the $\wppo,\wppp$ decay modes. The
particular features of this decay arising from the
application of chiral perturbation and vector
meson dominance \cite{paul1,bramon,bramon1} were
supplemented here by the inclusion of
$\omega-\rho$ mixing. Taking all these
contributions into account, we predict
$\Gamma(\wppo)=(390\pm96){\rm eV},$ which is
smaller though barely consistent with the existing
experimental value \cite{alde} of $\Gamma^{\rm
(exp)} =610\pm230 {\rm eV}.$ The measurements of
both channels and of their spectra, will afford
also the detection of the $\alpha^3$ effect which
we described and will allow to determine whether a
serious discrepancy between theory and experiment
occurs in this case.

The research of P.S. has been supported in part by
the Fund for Promotion of Research at the
Technion.

%\begin{figure}[b]
%\epsfxsize=8 truecm
%\epsfysize=4  truecm

%\moveright1in\hbox{
%\epsffile{anom.eps}}

%\caption{\baselineskip 16pt
%The anomaly graph for
%$B^*(D^*)\rightarrow B(D)
%\gamma\gamma.$
%}
%\label{fig:2}
%\end{figure}

\begin{figure}[b]
\epsfxsize=10 truecm
\epsfysize=4 truecm

\moveright1in\hbox{
\epsffile{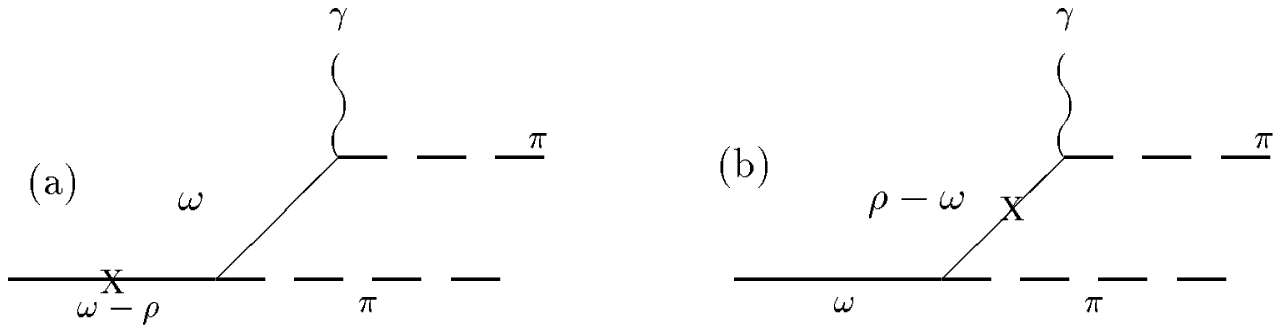}}

\caption{\baselineskip 16pt
\wppo decay via mixing: (a) in the wave function
(b) in the propagator.}
\label{fig:2}
\end{figure}

\end{document}